\documentclass[prl,aps,twocolumn,groupaddress]{revtex4}
\usepackage{graphicx,epsfig}
\usepackage{dcolumn,array}
\usepackage{bm}
\usepackage{amsmath}
\usepackage{color}
\usepackage{subfigure}
\usepackage{times}
\usepackage{amsfonts,amssymb}
\usepackage{tikz}
\usepackage{pgfplots}
\usepackage{verbatim}
\usepackage[colorlinks=false]{hyperref}

\begin{document}
\title{ $\ell_1$-norm in three-qubit quantum entanglement constrained by Yang-Baxter equation}

\author{Li-Wei Yu}
\email{nkyulw@yahoo.com}
 \affiliation{Theoretical Physics Division, Chern Institute of Mathematics, Nankai University, Tianjin 300071, China}
\author{Mo-Lin Ge}
\email{geml@nankai.edu.cn}
 \affiliation{Theoretical Physics Division, Chern Institute of Mathematics, Nankai University, Tianjin 300071, China}



\date{\today}


\begin{abstract}
{Usually the $\ell_2$-norm plays vital roles in quantum physics, acting as the probability of states.  In this paper, we show the important roles of $\ell_1$-norm in Yang-Baxter quantum system, in connection with both the braid matrix and quantum entanglements. Concretely,  we choose the 2-body and 3-body S-matrices, constrained by Yang-Baxter equation.  It has been shown that for 2-body case, the extreme values of $\ell_1$-norm lead to two types of braid matrices and 2-qubit Bell states. Here we show that  for the 3-body case, due to the constraint of YBE, the extreme values of $\ell_1$-norm lead to both 3-qubit $|GHZ\rangle$ (local maximum) and $|W\rangle$ (local minimum) states, which cover all  3-qubit genuine entanglements for pure  states under SLOCC. This is  a more convincing proof for the roles of $\ell_1$-norm in quantum mechanics.}

\end{abstract}


\maketitle

\noindent{\em Introduction}--The role of $\ell_1$-norm in information theory has been widely and well studied due to its importance, especially in Compressed Sensing theory~\cite{donoho2006compressed,candes2006near}. The minimum value of $\ell_1$-norm is utilized for quantifying sparsity of signals, under which condition the signal can be reconstructed more quickly. On the other hand, in quantum mechanics, the $\ell_2$-norm usually plays vital roles, e.g. as the probability in describing the overlap between two different quantum states.  For a given state $|\psi\rangle$ spanned by eigenstates $\{a_n\}$ of the system, $|\psi\rangle=\sum_{i}c_i|a_i\rangle$, the $\ell_2$-norm can be  defined by $\|\psi\|_{\ell_2}=\sum_i|c_i|^2$ for normalization. Almost all of the quantum physical consequences one can obtain are defined in terms of  $\ell_2$-norm. 

Since the $\ell_1$-norm has been successfully utilized in classical information theory,  it is interesting to guess, whether does the $\ell_1$-norm also apply in quantum information theory, or in quantum mechanics? Here similar to the definition of $\ell_2$-norm, the $\ell_1$-norm for a quantum state $|\psi\rangle=\sum_{i}c_i|a_i\rangle$ can be defined as follows
\begin{equation}
\|\psi\|_{\ell_1}=\sum_i|c_i|.
\end{equation}
In recent years, people are becoming aware of   $\ell_1$-norm in quantum mechanics and finding its physical interpretations. Some progress, though rarely, have been made, 
 such as in quantum process tomography \cite{kosut2008quantum}, Yang-Baxter equation(YBE) associated with Wigner $D^j$-functions \cite{niu2011role,ge2016yang,yu2016z3} and quantifying coherence \cite{baumgratz2014quantifying}, {\em et al.} 

Here we are interested in the $\ell_1$-norm in quantum information through  the solutions $\breve{R}$ of YBE.  The motivations for choosing solutions $\breve{R}$ of YBE are:
 \begin{enumerate}
 \item $\breve{R}(\theta)$ is closely related to concepts in quantum properties, such as quantum entanglement, topological quantum computation~\cite{nayak2008non,wang2010topological} and so on.  The YBE solution $\breve{R}$ can be obtained via parametrizing Bell braid matrix, which generates 2-qubit (even $N$-qubit) maximal entanglement\cite{kauffman2004braiding,Yu2018}. The approach of parametrizing braid matrix here is also named Yang-Baxterization~\cite{Jimbo1986quantum,jones1989certain,ge1991explicit,cheng1991yang,ge1993trigonometric}. Hence $\breve{R}$ can be connected to both 2-qubit quantum entanglement and anyonic braiding system associated to topological quantum computation.  
 
  \item $\breve{R}$ can be regarded as a 2-body S-matrix~\cite{zamolodchikov1979factorized}, and can reasonably compose 3-body S-matrix (till $N$-body) constrained by YBE.  Hence the results one obtains for 2-body may be extended to the many-body case. 
 \end{enumerate}

 In the previous works associated with YBE\cite{niu2011role,ge2016yang}, the authors have shown the motivation of applying $\ell_1$-norm in Wigner $D^{j}$-function to obtain 2-body anyonic physical models associated with $SU(2)_2$ Chern Simons theories. By defining the $\ell_1$-norm of $D^{\frac{1}{2}}(\theta,\varphi)$, the local maximal $\|D^\frac{1}{2}\|_{\ell_1}$ leads to Majorana braid representation (type-II)~\cite{ivanov2001non} and maximal 2-qubit entanglement, while the local minimal $\|D^\frac{1}{2}\|_{\ell_1}$ leads to the type-I braid representation associated with permutation. Furthermore, the results are extended to the 2-body parafermionic solution $\breve{R}$ of YBE\cite{yu2016z3}.

In this letter, we show that $\ell_1$-norm not only applies  to the 2-body system, but also in the 3-body system constrained by YBE, especially in obtaining the 3-qubit genuine entanglement. As an extension of  the maximal $\ell_1$-norm leading to 2-qubit maximal entanglement, the extreme values of $\ell_1$-norm for $\breve{S}_{123}$ leads to both 3-qubit $|GHZ\rangle$(local maximum) and $|W\rangle$ (local minimum) states, which are the only two types of genuine entanglement under stochastic local operation and classical communication (SLOCC)\cite{dur2000three}.


\bigskip
\noindent {\em Review of $\ell_1$-norm in 2-body Yang-Baxter system}--Here for completeness, we would give a brief review about $\ell_1$-norm in 2-qubit system. We start from the YBE, reads~\cite{yang1967some,yang1968matrix,baxter1972partition}
\begin{equation}
  \breve{R}_{12}(\theta_1) \breve{R}_{23}(\theta_2) \breve{R}_{12}(\theta_3)=\breve{R}_{23}(\theta_3) \breve{R}_{12}(\theta_2) \breve{R}_{23}(\theta_1),
 \end{equation}
 where $\breve{R}_{12}(\theta)=\breve{R}(\theta)\otimes I$, $\breve{R}_{23}(\theta)=I \otimes\breve{R}(\theta)$, $I$ represents 2D identity matrix, $\breve{R}$ is the $4\times4$ matrix located in 2-qubit space.
 
 The type-II solution of YBE comes from the parametrization of Bell braid matrix~\cite{dye2003unitary,kauffman2004braiding}
 \begin{equation}\label{typeIIbm}
 B=\exp[i\frac{\pi}{4}\sigma^y\otimes\sigma^x],
 \end{equation}
 and reads~\cite{chen2007braiding,chen2008berry}
 \begin{equation}\label{typeIIrm}
 \breve{R}(\theta)=\exp[i\theta\sigma^y\otimes\sigma^x].
 \end{equation}
 Applying $\breve{R}(\theta)$ on natural basis $|00\rangle$, it leads to~\cite{chen2007braiding} 
 \begin{equation}\label{xi}
|\xi\rangle= \breve{R}(\theta)|00\rangle=\cos\theta|00\rangle+\sin\theta|11\rangle.
 \end{equation}
 Hence $\theta$ describes the continuous entangled degree for 2-qubit pure states. When $\theta=\frac{\pi}{4}$, the state is Bell state with the maximal entangled degree and the $\breve{R}(\theta=\frac{\pi}{4})$ reduces to braid matrix $B$, which obeys braid relation $B_{12}B_{23}B_{12}=B_{23}B_{12}B_{23}$ for $B_{12}=B\otimes I$, $B_{23}=I\otimes B$. Since this braid matrix $B$ leads to the Bell state, it is also called Bell braid matrix.
 
\begin{figure}[h]
\includegraphics[width=20pc]{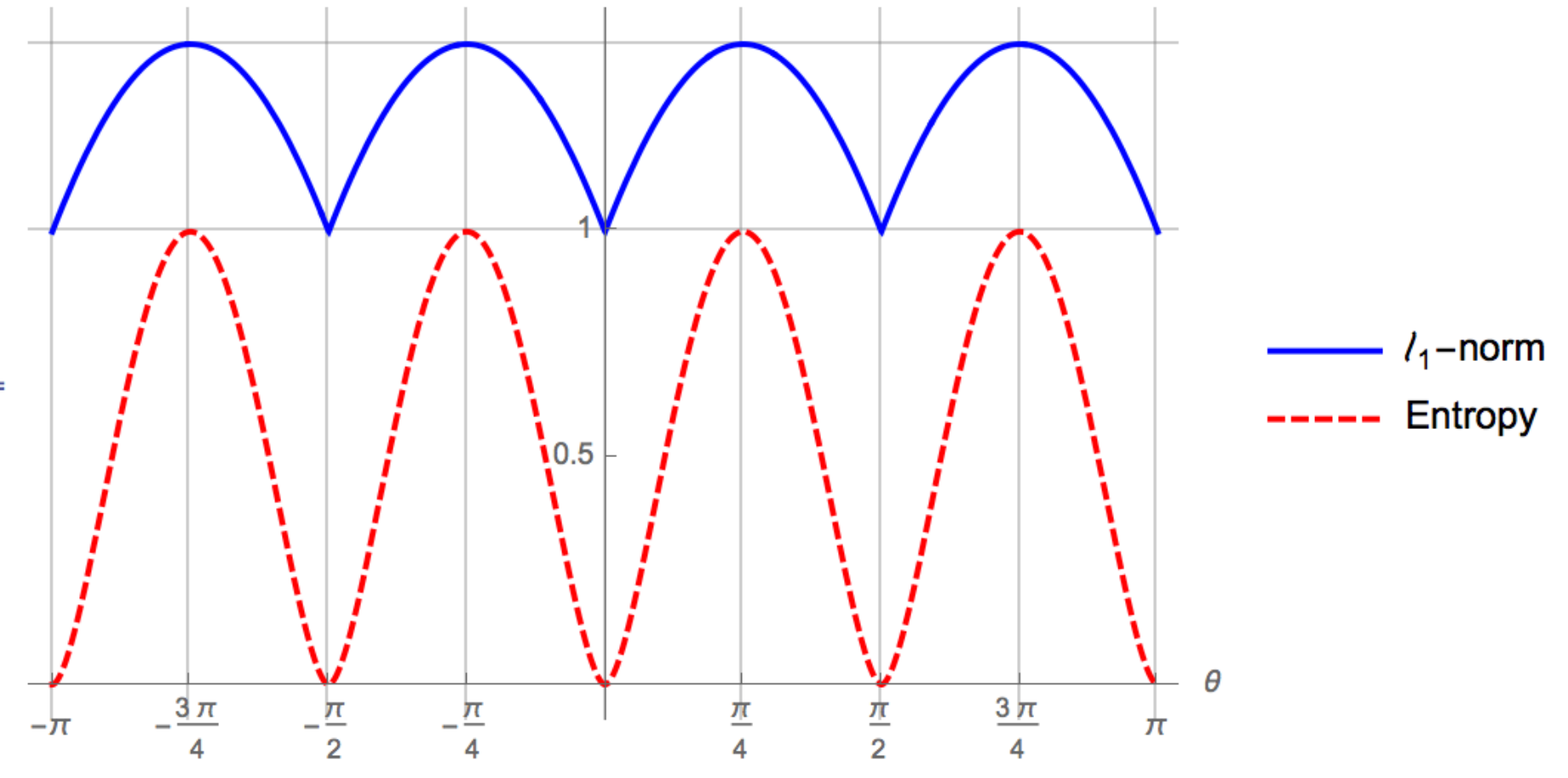}\hspace{2pc}%
\caption{\label{2L1vN}Von Neumann entropy and $\ell_1$-norm of $D^{\frac{1}{2}}$-function and the corresponding von Neumann entropy for $|\xi\rangle$ as functions of $\theta$. The von Neumann entropy is labeled by the red dashed line, and the $\ell_1$-norm is labeled by blue solid line. Both of them achieve the local maximum at $\theta=\frac{\pi}{4}$, at which value the $\breve{R}(\theta)$ reduces Bell braid matrix.}
\end{figure}

On the other hand, we introduce the  Wigner $D^j(\theta,\varphi)$-function that satisfies the braid relation~\cite{benvegnu2006uncertainty}
\begin{equation}
D^j(\theta,0)D^j(\theta,\phi)D^j(\theta,0)=D^j(\theta,\phi)D^j(\theta,0)D^j(\theta,\phi),
\end{equation}
with the constraint (for all $j$)
\begin{equation}
\cos\phi=\frac{\cos2\theta}{1-\cos2\theta}.
\end{equation}
Yang-Baxterization of the braid relation leads to YBE~\cite{niu2011role},
\begin{equation}
D^j(\theta_1,0)D^j(\theta_2,\phi)D^j(\theta_3,0)=D^j(\theta_3,\phi)D^j(\theta_2,0)D^j(\theta_1,\phi),
\end{equation}
with the constraint (for all $j$)
\begin{equation}\label{p123}
\cos\phi=\frac{1}{2}\left[\frac{(\tan\theta_1+\tan\theta_3)-\tan\theta_2}{\tan\theta_1\tan\theta_2\tan\theta_3}-1\right].
\end{equation}
Let us now define the $\ell_1$-norm of the $D^j$-function. The $\|D^j\|_{\ell_1}$ can be defined as
\begin{equation}\label{WignerL1}
\|D^j\|_{\ell_1}=\frac{1}{2j+1}\sum_{m,n}^{2j+1}|D^j_{mn}|.
\end{equation}
Taking $D^{\frac{1}{2}}(\theta,\phi)=\left[\begin{matrix}\cos\theta & -\sin\theta e^{-i\phi}\\ \sin\theta e^{i\phi}&\cos\theta\end{matrix}\right]$ as example, we have
\begin{equation}
\|D^{\frac{1}{2}}(\theta,\phi)\|_{\ell_1}=|\cos\theta|+|\sin\theta|.
\end{equation}
In the previous papers~\cite{niu2011role,GE2012,Ge2014Yangbaxter}, it has been shown that extremization of $\|D^{\frac{1}{2}}\|_{\ell_1}$ leads to the following two cases, see Fig.~\ref{2L1vN}.
\begin{itemize}
\item Local minimum  $\|D^{\frac{1}{2}}(\theta,\phi)\|_{\ell_1}$, $\theta=\frac{\pi}{2}$, $\phi=\frac{2\pi}{3}$. In this case, the 2D braid matrix is called Type-I. It also corresponds to the $4\times4$ permutation braid matrix(see Appendix A in~\cite{supp}). The parametrized matrix is the simplest rational solution of YBE proposed by Yang~\cite{yang1967some}(we call it type-I solution),  obeying Galilean additivity $\tan\theta_2=\tan\theta_1+\tan\theta_3$. It is the traditional 6-vertex model and corresponds to the integrable models.
\item Local maximum $\|D^{\frac{1}{2}}(\theta,\phi)\|_{\ell_1}$, $\theta=\frac{\pi}{4}$, $\phi=\frac{\pi}{2}$. In this case, the 2D braid matrix is called type-II. This is nothing but the Ising braid matrix in 4-anyon fusion sparse encoding space~\cite{Bravyi2006}. It corresponds to the $4\times4$ Bell matrix proposed in Eq. (\ref{typeIIbm})(see Appendix A in~\cite{supp}). The parametrized YBE obeys Lorentzian additivity $\tan\theta_2=\frac{\tan\theta_1+\tan\theta_3}{1+\tan\theta_1\tan\theta_3}$. Such a local maximum coincides with the maximum of von Neumann entropy for the state $|\xi\rangle$ in Eq.~(\ref{xi}) at $\theta=\frac{\pi}{4}$.
\end{itemize}
The $2\times 2$ and $4\times4$ solutions are connected by topological bases, for details see Appendix A in~\cite{supp}. In quantum computation,  $4\times4$ type-I braid matrix is a swapping gate, while the $4\times4$ type-II braid matrix is an entangling gate. Hence the two braid matrices are totally different from each others. 

In summary, the extremums of $\ell_1$-norm lead to two different solutions of YBE. Especially, the local maximum of $\|D^{\frac{1}{2}}\|_{\ell_1}$ is pivotal in leading to Ising braiding and the maximal 2-qubit entanglement among all the possibilities. 


%

\bigskip
 
 \noindent {\em $\ell_1$-norm in 3-body Yang-Baxter system} -- YBE allows the factorization of 3-body S-matrix into 2-body S-matrices(see Appendix C in~\cite{supp}). This is our motivation for choosing the Yang-Baxter system.  Under the constraint of YBE,  one can construct the 3-body S-matrix from the type-II $\breve{R}$-matrix(see Appendix C in~\cite{supp}), as\cite{yu2014factorized}
\begin{equation}\label{3MFExp}
  \breve{S}_{123}(\eta,\beta)=e^{\eta \left(\vec{n}\cdot\vec{\Lambda}\right)},
\end{equation}
where 
\begin{eqnarray*}
 \cos\eta &=& \cos\theta_2 \cos\left(\theta_1+\theta_3\right),\\
 \sin\eta &=& \sin\theta_2\sqrt{1+\cos^2(\theta_1-\theta_3)},\\
 \vec{n} &=& \left(
  \begin{array}{ccc}
  \tfrac{1}{\sqrt{2}}\cos\beta,& \tfrac{1}{\sqrt{2}}\cos\beta,& \sin\beta
  \end{array}\right),\\
  \vec{\Lambda} 
  		&=&(-i\sigma^y\sigma^x I,\, -iI\sigma^y\sigma^x,\, -i\sigma^y\sigma^z\sigma^x),\\
  \cos\beta &=& \tfrac{\sqrt{2}\cos\left(\theta_1-\theta_3\right)}{\sqrt{1+\cos^2\left(\theta_1-\theta_3\right)}},\\               
  \sin\beta &=& \tfrac{-\sin\left(\theta_1-\theta_3\right)}{\sqrt{1+\cos^2\left(\theta_1-\theta_3\right)}}.
 \end{eqnarray*}
Here there are only two free parameters $\eta$ and $\beta$  due to the constraint of YBE. And it is easy to check that the three segments in $\vec{\Lambda}$ obey $su(2)$ algebra. 

Following the type-II Yang-Baxter solution $\breve{R}(\theta)$(Eq. \ref{typeIIrm}) in describing 2-qubit entanglement, we find that the factorized 3-body S-matrix $\breve{S}_{123}(\eta,\beta)$ also generates 3-qubit entanglement. 
Applying  $\breve{S}_{123}$ on the state $|000\rangle$, one obtains
\begin{equation}
\begin{split}
&|\Psi\rangle=\breve{S}_{123}|000\rangle\\
&=\cos\eta|000\rangle-\frac{\cos\beta\sin\eta}{\sqrt{2}}\left(|011\right\rangle+\left|110\right\rangle)-\sin\beta\sin\eta|101\rangle.
\end{split}
\end{equation}

 
 Significantly, $\breve{S}_{123}(\eta,\beta)$ is able to generate the two types of maximal 3-qubit entangled states, say, GHZ state and $W$ state.
\begin{itemize}
\item When $\eta=\frac{\pi}{3}, \, \beta=\textrm{arccot} \sqrt{2}$, 
\begin{equation}
\begin{aligned}
\breve{S}_{123}|000\rangle&=\frac{1}{2}(|000\rangle+|011\rangle+|101\rangle+|110\rangle)\\
&=\frac{1}{\sqrt{2}}(|+\rangle^{\otimes3}+|-\rangle^{\otimes3}), \,\textrm{(GHZ state)}.
\end{aligned}
\end{equation}
\item When $\eta=\frac{\pi}{2}, \, \beta=\textrm{arccot} \sqrt{2}$, 
\begin{equation}
\breve{S}_{123}|000\rangle=\frac{1}{\sqrt{3}}(|011\rangle+|101\rangle+|110\rangle),\,\textrm{(W state)}.
\end{equation}
\end{itemize} 
It is well known that in 3-qubit pure state, under SLOCC, GHZ state and $W$ state are the only two types of non-equivalent genuine entangled states\cite{dur2000three}. Hence our 3-body $S$-matrix describes all 3-qubit genuine entangled pure states in this sense.

Here we claim that the $\ell_1$-norm plays the pivotal role in screening out the genuine entangled states. Let us now define the $\ell_1$-norm of $\breve{S}_{123}$, which is shown equivalent to $\|\Psi\|_{\ell_1}$, as
\begin{equation}\label{S3L1}
\begin{split}
&\|\breve{S}_{123}\|_{\ell_1}=\|\Psi\|_{\ell_1}=|\cos\eta|+\sqrt{2}|\cos\beta\sin\eta|+|\sin\beta\sin\eta|.
\end{split}
\end{equation}

\begin{figure}[ht]
\includegraphics[width=20pc]{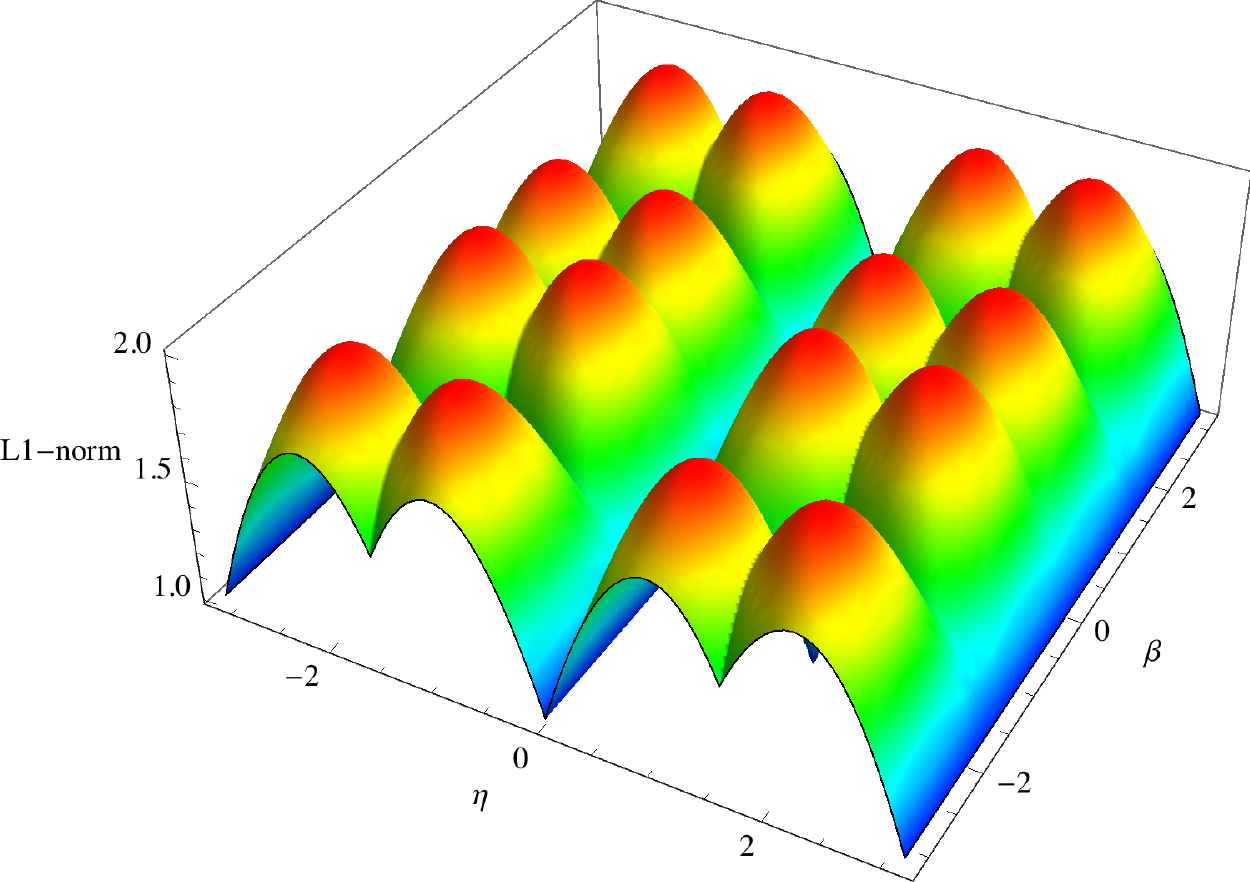}\hspace{2pc}%
\caption{\label{3DL1etabeta} $\ell_1$-norm of $|\Psi\rangle$ and $\breve{S}_{123}$ as a function of $\eta$ and $\beta$.  Both $|GHZ\rangle$ and $|W\rangle$ locate on the extremum points.}
\end{figure}
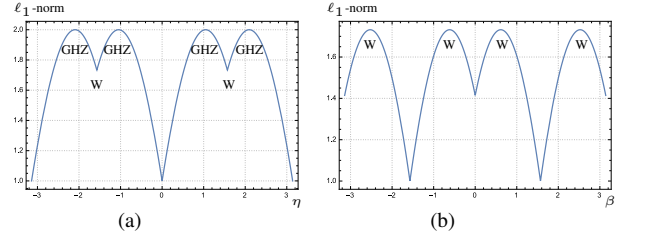
\begin{figure}
\subfigure[]{\label{Figa}
  \begin{picture}(90,80)
    \put(0,0){\includegraphics[width=9pc]{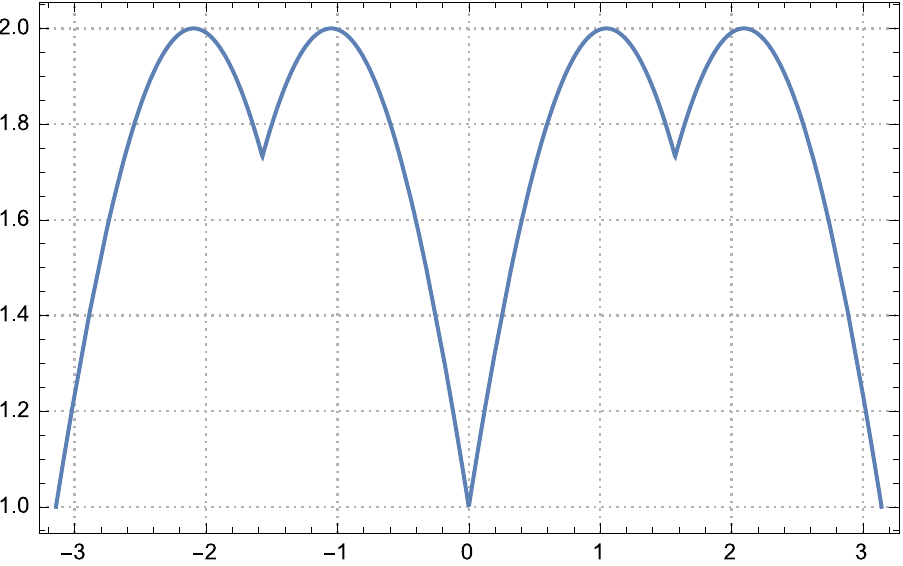}}
    \put(79,42){{\tiny W}}
    \put(29,42){{\tiny W}} 
    \put(18,55){{\tiny GHZ}}
    \put(34,55){{\tiny GHZ}}
    \put(67,55){{\tiny GHZ}}
    \put(84,55){{\tiny GHZ}}
    \put(105,-2){{\tiny $\eta$}}
    \put(0,71){{\tiny $\ell_1$-norm}}
  \end{picture}}
  \hspace{0.3in}
  \subfigure[]{\label{Figb}
  \begin{picture}(90,80)
    \put(0,0){\includegraphics[width=9pc]{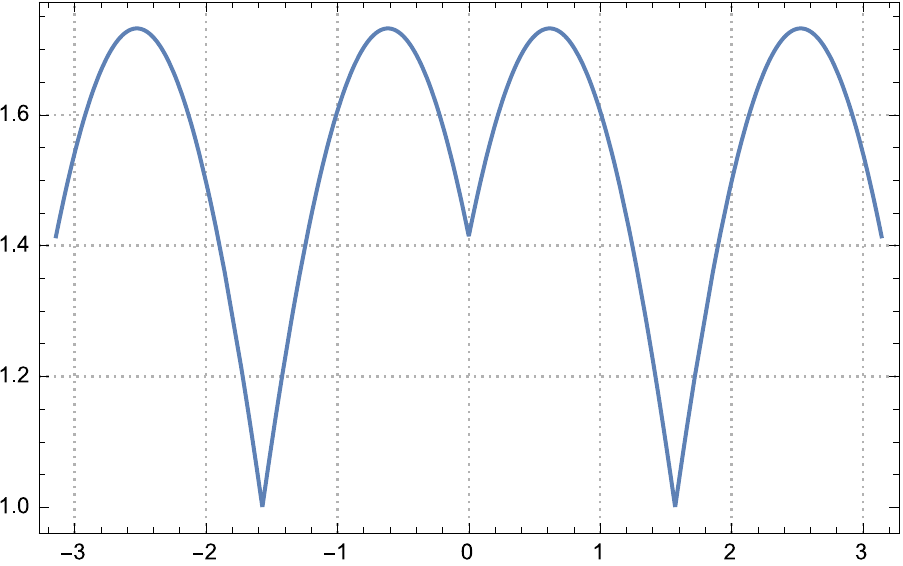}}
    \put(14,57){{\tiny W}}
    \put(44,57){{\tiny W}}
    \put(64,57){{\tiny W}}
    \put(94,57){{\tiny W}}
        \put(105,-2){{\tiny $\beta$}}
    \put(0,71){{\tiny $\ell_1$-norm}}
  \end{picture}}
\caption{\label{3-bodyL1-normsec1}(a) {\bf The section $\beta=\textrm{arccot} \sqrt{2}$.} $\ell_1$-norm of $|\Psi\rangle$ and $\breve{S}_{123}$ as a function of $\eta$. $|GHZ\rangle$ on local maximal points; $|W\rangle$ on local minimal points.\\ (b) {\bf The section $\eta=\pi/2$.} $\ell_1$-norm of $|\Psi\rangle$ and $\breve{S}_{123}$ as a function of $\beta$. $|W\rangle$ on local maximal points. }
\end{figure}

We can see from Fig.~\ref{3DL1etabeta} that when $\ell_1$-norm arrives at the ``mountaintop'', i.e. the local maximum, the parameters $\{\eta=\frac{\pi}{3}, \,\beta=\textrm{arccot}\sqrt{2}\}$ correspond to the 3-qubit $GHZ$ state. While the $\ell_1$-norm arrives at the ``saddle point'', the parameters $\{\eta=\frac{\pi}{2}, \,\beta=\textrm{arccot}\sqrt{2}\}$ correspond to the 3-qubit $W$ state.  The saddle point is much strange. It is the local maximal point along the section $\eta=\frac{\pi}{2}$ with varying $\beta$, while the local minimal point along the section $\beta=\textrm{arccot} \sqrt{2}$ with varying $\eta$. The ``saddle point''  corresponding to $|W\rangle$ can be observed from two sections of Fig.~\ref{3DL1etabeta}.  See Fig.~\ref{3-bodyL1-normsec1}.  When $\beta=\textrm{arccot} \sqrt{2}$, the section holds both $GHZ$ and $W$ states locating on the extreme points, while for the section $\eta=\pi/2$, the local maximum points  correspond to the $W$ states.  

\bigskip
 
 \noindent {\em 3-body $S$-matrix in $D^{1/2}$-form} -- In above section, we discuss the 3-body $S$-matrix in tensor product space, where each ``body'' represents one qubit lattice. Then the 3-body $S$-matrix $\breve{S}_{123}$ describes the properties of the three interacting qubit lattices. Indeed, the  $\breve{S}_{123}$ can be reduced to a lower 2D matrix under the two 4-strand topological bases $\{|e_1\rangle\, |e_2\rangle\}$, graphically (see Appendix A2 in~\cite{supp} for details)
  \begin{equation}\label{topological basis}
\begin{split}
   &|e_1\rangle =  \frac{1}{d}\,\, \begin{tikzpicture}[baseline=-\dimexpr\fontdimen22\textfont2\relax]
\draw[gray,line width=1.0pt] (-0.1,-0.1)--(-0.1,0.4) (-0.7,-0.1)--(-0.7,0.4) (0.1,-0.1)--(0.1,0.4) (0.7,-0.1)--(0.7,0.4);
\draw[gray,line width=1.0pt] (-0.7,-0.1) arc(180:360:0.3) (0.1,-0.1) arc(180:360:0.3);
\end{tikzpicture}\, , \quad (d=\sqrt{2})\\
   &|e_2\rangle = \frac{1}{\sqrt{d^2-1}}\left(\,\, \begin{tikzpicture}[baseline=-\dimexpr\fontdimen22\textfont2\relax]
\draw[gray,line width=1.0pt] (-0.3,-0.2)--(-0.3,0.4) (-0.7,-0.3)--(-0.7,0.4) (0.3,-0.2)--(0.3,0.4) (0.7,-0.3)--(0.7,0.4);
\draw[gray,line width=1.0pt] (-0.3,-0.19) arc(240:300:0.6) (-0.7,-0.29) arc(240:300:1.4);
\end{tikzpicture} \,\,  - \frac{1}{d} \,\, \begin{tikzpicture}[baseline=-\dimexpr\fontdimen22\textfont2\relax]
\draw[gray,line width=1.0pt] (-0.1,-0.1)--(-0.1,0.4) (-0.7,-0.1)--(-0.7,0.4) (0.1,-0.1)--(0.1,0.4) (0.7,-0.1)--(0.7,0.4);
\draw[gray,line width=1.0pt] (-0.7,-0.1) arc(180:360:0.3) (0.1,-0.1) arc(180:360:0.3);
\end{tikzpicture}\,\, \right),
\end{split}
\end{equation}
where each strand represents an Ising anyon with quantum dimension $d=\sqrt{2}$.

 We denote the reduced 2D $S$-matrix by $\breve{S}'_{123}$. Under bases $\{|e_1\rangle,\,|e_2\rangle\}$,  $\breve{S}'_{123}$ takes the following form(see Appendix C in~\cite{supp})
 \begin{equation}
 \breve{S}'_{123}=\left[\begin{matrix}\cos\eta+\frac{i}{\sqrt{2}}\cos\beta\sin\eta, &  (\frac{i}{\sqrt{2}}\cos\beta+\sin\beta)\sin\eta \\ (\frac{i}{\sqrt{2}}\cos\beta-\sin\beta)\sin\eta, & \cos\eta-\frac{i}{\sqrt{2}}\cos\beta\sin\eta\end{matrix}\right].
 \end{equation}
 Physically speaking, this 3-body $S$-matrix $\breve{S}'_{123}$ is  composed from the Wigner $D^{1/2}$-functions(2-anyon S-matrix), and describes the interactions between the first three strands, i.e. the three Ising anyons. We call $\breve{S}'_{123}$ the 3-anyon $S$-matrix. Following the definition of $\ell_1$-norm for Wigner $D$-function in Eq.~(\ref{WignerL1})\cite{niu2011role}, here we define the $\ell_1$-norm for $\breve{S}'_{123}$, as
 \begin{equation}
 \begin{split}
 \|\breve{S}'_{123}\|_{\ell_1}=&|\cos\eta+\frac{i}{\sqrt{2}}\cos\beta\sin\eta|\\
 &+|\frac{i}{\sqrt{2}}\cos\beta\sin\eta+\sin\beta\sin\eta|,\\
 =&|\cos\eta|+\sqrt{2}|\cos\beta\sin\eta|+|\sin\beta\sin\eta|.
 \end{split}
 \end{equation}
In comparison with $\|\breve{S}_{123}\|_{\ell_1}$ in Eq.~(\ref{S3L1}), we find that $\|\breve{S}'_{123}\|_{\ell_1}=\|\breve{S}_{123}\|_{\ell_1}$. Therefore, although the two S-matrices describe two different types of 3-body interactions,  the $\ell_1$-norms are invariant. 

To gain more intuitive physical inferences,  let us make a comparison between the $\ell_1$-norm and  von Neumann entropy of the 3-body $S$-matrix.  The von Neumann entropy $\|\breve{S}'_{123}\|_{E}$ of $\breve{S}'_{123}$ can be defined  via $|\psi\rangle=\breve{S}'_{123}\left(|e_1\rangle,\, |e_2\rangle\right)^{\rm T}$, denoted by $\|\psi\|_{E}$,
\begin{equation}
\begin{split}
&\|\breve{S}'_{123}\|_{E}=\|\psi\|_{E}\\
=&-|\cos\eta+\frac{i}{\sqrt{2}}\cos\beta\sin\eta|^2\log_2|\cos\eta+\frac{i}{\sqrt{2}}\cos\beta\sin\eta|^2\\
&-|(\frac{i}{\sqrt{2}}\cos\beta+\sin\beta)\sin\eta|^2\log_2|(\frac{i}{\sqrt{2}}\cos\beta+\sin\beta)\sin\eta|^2.
\end{split}
\end{equation}

\begin{figure}[ht]
\includegraphics[width=20pc]{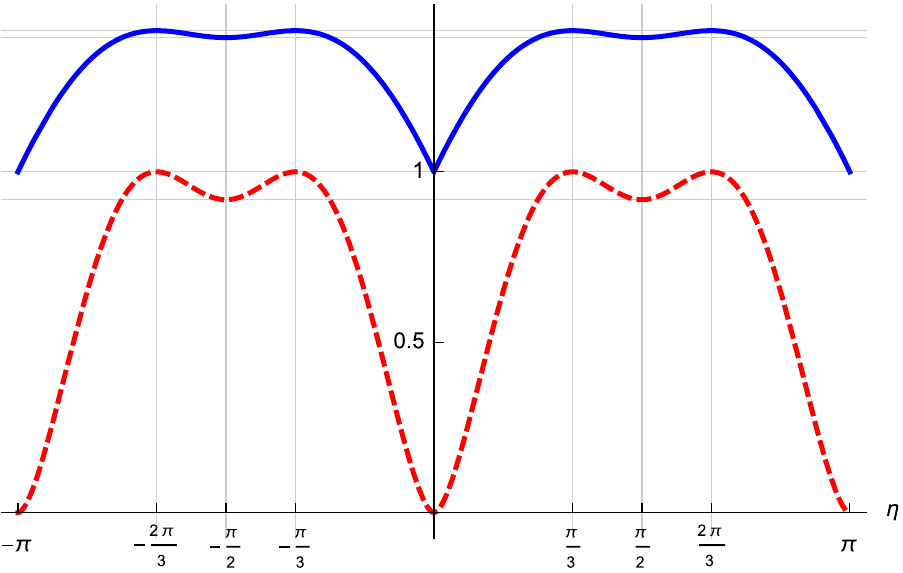}\hspace{2pc}%
\caption{\label{DF3BodyL1VN} Von Neumann entropy and $\ell_1$-norm of $\breve{S}'_{123}$ as a function of $\eta$, at the section $\beta=\textrm{arccot}\sqrt{2}$. The $\ell_1$-norm is labeled by blue solid line, while the von Neumann entropy is labeled by red dashed line.  $|GHZ\rangle$  locates on the local maximum and $|W\rangle$ locates on the local minimum.}
\end{figure}

For simplicity, we only plot a section at $\beta=\textrm{arccot}\sqrt{2}$. Fig.~\ref{DF3BodyL1VN} simply shows that the $\ell_1$-norm and von Neumann entropy achieve their local extremum at the same $\eta$, which leads to the 3-qubit genuine entanglement.   
 
 Hence the two representations: tensor product representation and Wigner $D^{1/2}$-function representation coincide with each others at the extremum of $\ell_1$-norms as well as the two types of 3-qubit genuine entanglement.
 
 \bigskip
 
 \noindent{\em Conclusion} -- In conclusion,  we have predicted the role of $\ell_1$-norm in 3-body quantum system  constrained by YBE. We show that the 3-body $S$-matrix $\breve{S}_{123}$ covers all 3-qubit genuine entanglement under SLOCC. Moreover, we define the $\ell_1$-norm of the 3-body $S$-matrix, and find that all the 3-qubit genuine entangled pure states(including GHZ and W) locate on the local extremums of the $\ell_1$-norm.  Furthermore, to check the consistency between two types of 3-body $S$-matrices in qubit tensor space and Ising anyon fusion space, we calculate the corresponding $\ell_1$-norm and von Neumann entropy. We make comparisons between the extremum of the above quantities and find all of them achieve the local extremum at the same domain values, which are similar to the 2-body case presented in~\cite{niu2011role,ge2016yang}. Hence the 3-body results are exactly extensions of the 2-body cases. Therefore, it is more convincing that the $\ell_1$-norm does play pivotal roles in this Yang-Baxter system as well as quantum entanglement.  
 
 Again, we emphasize that by introducing YBE to our system, the 3-body S-matrix can be reasonably decomposed into the 2-body S-matrices, which greatly reduces the hardness in dealing with 3-body problem. For example, in covering all 3-qubit genuine entanglement via $\breve{S}_{123}$, we only need two free parameters $\eta$ and $\beta$, whose freedom is far less than the freedom of a 3-qubit ($\mathbb{C}^8$) space. This is our motivation. 
 
In principle,  the N-body $S$-matrix can also be decomposed into 2-body $S$-matrices by means of YBE. Further,  we would discuss the $\ell_1$-norm in 4-body system even N-body.

\bigskip

\noindent{\em Acknowledgement} -- This work is in part supported by the National Natural Science Foundation of China (Grant No. 11475088). L-W, Yu is supported by China Postdoctoral Science Foundation (Grant No. 2018M641622).

\bibliographystyle{unsrt}

\pagebreak
\pagebreak
\clearpage
\setcounter{equation}{0}
\setcounter{figure}{0}

\renewcommand{\theparagraph}{\bf}
\renewcommand{\theequation}{A\arabic{equation}}
\onecolumngrid
\flushbottom

\section{Supplementary Material}

%




\appendix
\subsection{Appendix A: Two types of braid matrices and their Yang-Baxterizations}
Typically,  the local unitary representation(LUR) of N-strand braid group $\mathcal{B}_N$takes the following tensor form in space $(\mathbb{C}^k)^{\otimes N}$:
\begin{equation}
\begin{split}
&B_iB_{i+1}B_{i}=B_{i+1}B_{i}B_{i+1},\\
&B_iB_j=B_jB_i,\quad (|i-j|>1)\\
&B_i=I\otimes I\otimes\cdots\otimes \underset{i,i+1}{B}\otimes I\cdots I\otimes I,
\end{split}
\end{equation}
where $B_i$'s are generators, $B$ is a  $k^2\times k^2$ braid matrix locating on the $i$-th and $i+1$-th lattices.

Yang-Baxterization of braid matrix $B_i\rightarrow \breve{R}_i(x)$ leads to the Yang-Baxter equation(YBE),
\begin{equation}
\breve{R}_i(x)\breve{R}_{i+1}(xy)\breve{R}_i(y)=\breve{R}_{i+1}(y)\breve{R}_i(xy)\breve{R}_{i+1}(x).
\end{equation}
The braid matrix associated with different algebraic structures correspond to different Yang-Baxterization approaches, including rational, trigonometric, elliptic, and so on.  In this paper, the braid matrices are only associated to Temperley-Lieb(TL) algebra, hence we only need the simplest rational Yang-Baxterization.

\subsection{A1. Rational Yang-Baxterization of braid matrix associated to TL algebra}
The $TL_N$ algebra has $N-1$ generators $\{T_i\}$, and has the following structure
\begin{equation}
\begin{split}
&T_{\textrm{i}}^2=d T_{\textrm{i}},\quad \textrm{d: quantum dimension} \\
&T_{\textrm{i}}T_{\textrm{i}\pm1}T_{\textrm{i}}=T_{\textrm{i}},\\
&T_{\textrm{i}}T_{\textrm{j}}=T_{\textrm{j}}T_{\textrm{i}}, \quad |i-j|>1.
\end{split}
\end{equation}
The braid representation $\{B_i\}$ associated to TL algebra $\{T_i\}$ can be  expressed as follows 
\begin{equation}\label{BTrelation}
B_i=\alpha  I +\alpha^{-1} T_i,\quad d=-\alpha^2-\alpha^{-2}\, {\rm (quantum\, dimension)}.
\end{equation}
 Usually, $\alpha$ is root of unity. 
 
 To obtain the Yang-Baxter equation,
\begin{equation}\label{YBEmu}
\breve{R}_{\textrm{i}}(\mu) \breve{R}_{\textrm{i}+1}(f(\mu, \nu)) \breve{R}_{\textrm{i}}(\nu)=\breve{R}_{\textrm{i}+1}(\nu) \breve{R}_{\textrm{i}}(f(\mu, \nu)) \breve{R}_{\textrm{i}+1}(\mu),
\end{equation}
the rational Yang-Baxterization yields the following solution, as
\begin{eqnarray}
&&\breve{R}_{\textrm{i}}(\mu)=\rho(\mu)[1+G(\mu)T_{\textrm{i}}],\label{Rmu}\smallskip\\
&&G(\mu)=\frac{\mu}{a_0-d\mu/2},\label{Gmu}\\
&&f(\mu,\nu)=\frac{\mu+\nu}{1+\beta^2\mu\nu}, \, \beta^2=\frac{d^2-4}{(2a_0)^2}\label{fmu}.
\end{eqnarray}
Here  $T_{\textrm{i}}$ is Temperley-Lieb algebra(TLA) generator, $d$ represents the quantum dimension of TLA and $a_0$ is a free parameter. If we express $T_{\textrm{i}}$ in terms of the known braid operator $B_{\textrm{i}}$, then the solution of YBE can be obtained.

\subsection{A2. The type-I $4\times4$ braid matrix and Yang-Baxterization}
The type-I braid matrix is nothing but the 2-qubit permutation, which  reads
\begin{equation}
^1B=P=\left[
   \begin{array}{cccc}
   1 & 0 & \ 0 & \ 0\\
   0 & 0 & \ 1 & \ 0 \\
   0 & 1 & \ 0 & \ 0 \\
   0 & 0 & \ 0 & \ 1
   \end{array} \right],
\end{equation}

where $P$ is the $4\times4$ representation of permutation, $P^2=1$. The braid matrix $^1B$ has two different eigenvalues and  corresponds to the TL element $^1T$, for $\alpha=i$, and quantum dimension $d=-\alpha^2-\alpha^{-2}=2$, (up to an over all phase $-i$ in comparison with  Eq.~(\ref{BTrelation}))
\begin{equation}
^1B=-i(\alpha I+\alpha^{-1} \,^1T)=I-\,^1T=P.
\end{equation}
The $^1T$-matrix corresponding T-L generators $^1T_i=I\otimes\cdots \underset{i,i+1}{^1T}\cdots\otimes I$ reads
\begin{equation}\label{TLB1}
^1T= \left[
   \begin{array}{cccc}
   0 & 0 & \ 0 & \ 0\\
   0 & 1 & \ -1 & \ 0 \\
   0 & -1 & \ 1 & \ 0 \\
  0 & 0 & \ 0 & \ 0
   \end{array} \right].
\end{equation}
with quantum dimension is $d=2$.
Substituting Eq.~(\ref{TLB1}) into Eq.~(\ref{Rmu}), one obtains the Yang-Baxterized matrix, denoted as $^1\breve{R}(u)$ (suppose $a_0=-1$)
\begin{equation}
^1\breve{R}(\mu)=\rho(\mu)\left(I+\frac{\mu}{a_0-d\mu/2}T\right)=\rho(\mu)\left(\frac{a_0 I-\mu I+\mu\, ^1T}{a_0-d\mu/2}\right)=\frac{1}{|\sqrt{1-\mu^2}|}(I-\mu (I-\,^1T))=\frac{1}{|\sqrt{1-\mu^2}|}(I+\mu P).
\end{equation}
To satisfy Yang-Baxter equation, the parameter relation in Eq. ~(\ref{fmu})  is shown to be Galilean: 
\begin{equation}
f(\mu,\nu)=\mu+\nu.
\end{equation}

\bigskip
\noindent {\bf Reducing type-I tensor YBE to lower dimension.}  $^1T_i$ can be expressed by spin-1/2 lattice sites
\begin{equation}
^1T_i=2|\phi_{i,i+1}\rangle\langle\phi_{i,i+1}|,
\end{equation}
where $|\phi_{i,i+1}\rangle=\frac{1}{\sqrt{2}}\left(|01\rangle-|10\rangle\right)$ represents  Bell state on $i$-th and $(i+1)$-th lattices.
In this case, one can also introduce two basis $|e_1\rangle$ and $|e_2\rangle$ represented by 4 spin-1/2 lattices for 4-strand T-L algebra,
\begin{align}
&|e_1\rangle=|\phi_{12}\rangle|\phi_{34}\rangle,\\
&|e_2\rangle=\frac{1}{\sqrt{3}}\left(2|\phi_{41}\rangle|\phi_{23}\rangle-|\phi_{12}\rangle|\phi_{34}\rangle\right),\\
&|\phi_{i,i+1}\rangle=\frac{1}{\sqrt{2}}\left(|\underset{i}{0}\underset{j}{1}\rangle-|\underset{i}{1}\underset{j}{0}\rangle\right).
\end{align}
Graphically, 
 \begin{equation}\label{topological basis1}
\begin{split}
   &|e_1\rangle =  \frac{1}{d}\,\, \begin{tikzpicture}[baseline=-\dimexpr\fontdimen22\textfont2\relax]
\draw[gray,line width=1.0pt] (-0.1,-0.1)--(-0.1,0.4) (-0.7,-0.1)--(-0.7,0.4) (0.1,-0.1)--(0.1,0.4) (0.7,-0.1)--(0.7,0.4);
\draw[gray,line width=1.0pt] (-0.7,-0.1) arc(180:360:0.3) (0.1,-0.1) arc(180:360:0.3);
\end{tikzpicture}\, , \quad (d=2)\\
   &|e_2\rangle = \frac{1}{\sqrt{d^2-1}}\left(\,\, \begin{tikzpicture}[baseline=-\dimexpr\fontdimen22\textfont2\relax]
\draw[gray,line width=1.0pt] (-0.3,-0.2)--(-0.3,0.4) (-0.7,-0.3)--(-0.7,0.4) (0.3,-0.2)--(0.3,0.4) (0.7,-0.3)--(0.7,0.4);
\draw[gray,line width=1.0pt] (-0.3,-0.19) arc(240:300:0.6) (-0.7,-0.29) arc(240:300:1.4);
\end{tikzpicture} \,\,  - \frac{1}{d} \,\, \begin{tikzpicture}[baseline=-\dimexpr\fontdimen22\textfont2\relax] 
\draw[gray,line width=1.0pt] (-0.1,-0.1)--(-0.1,0.4) (-0.7,-0.1)--(-0.7,0.4) (0.1,-0.1)--(0.1,0.4) (0.7,-0.1)--(0.7,0.4);
\draw[gray,line width=1.0pt] (-0.7,-0.1) arc(180:360:0.3) (0.1,-0.1) arc(180:360:0.3);
\end{tikzpicture}\,\, \right).
\end{split}
\end{equation}
Applying the T-L generators to the basis, one obtains
\begin{align}
&^1T_1|e_1\rangle=\,^1T_3|e_1\rangle=2|e_1\rangle,\quad ^1T_1|e_2\rangle=\,^1T_3|e_2\rangle=0,\\
&^1T_2|e_1\rangle=\frac{1}{2}\left(|e_1\rangle+\sqrt{3}|e_2\rangle\right),\quad ^1T_2|e_2\rangle=\frac{\sqrt{3}}{2}\left(|e_1\rangle+\sqrt{3}|e_2\rangle\right).
\end{align}
The corresponding $2\times2$ 4-strand braid matrices $^1B_i$ are($\gamma$ is arbitrary number)
\begin{equation}\label{Braid2}
^1B_1=\,^1B_3=\left[
   \begin{array}{cc}
   -1 & 0 \\
   0 & 1 
   \end{array} \right], \quad ^1B_2=\frac{1}{2}\left[
   \begin{array}{cc}
   1 & -\sqrt{3} \\
   -\sqrt{3} & -1 
   \end{array} \right].
\end{equation}

Then the  type-I $2\times2$ correspondence $^1\mathcal{R}_i$ reads
\begin{align}
&^1\mathcal{R}_1(\mu)= \, ^1\mathcal{R}_3(\mu)=\frac{1}{|\sqrt{1-\mu^2}|}\left[
   \begin{array}{cc}
  1-\mu & 0 \\
   0 & 1+\mu 
   \end{array} \right],\\
&^1\mathcal{R}_2(\mu)=\frac{1}{2|\sqrt{1-\mu^2}|}\left[
   \begin{array}{cc}
   2+\mu & -\sqrt{3}\mu \\
   -\sqrt{3}\mu &  2-\mu
   \end{array} \right].
\end{align}
They satisfy the YBE and the parameters obey the Galilean additivity,
\begin{equation}
^1\mathcal{R}_1(\mu) \,^1\mathcal{R}_2(\mu+\nu) \,^1\mathcal{R}_1(\nu)= \,^1\mathcal{R}_2(\nu) \,^1\mathcal{R}_1(\mu+\nu)\,^1\mathcal{R}_2(\mu). 
\end{equation}

\subsection{A3. The type-II $4\times4$ braid matrix and Yang-Baxterization}

The type-II braid matrix is the Bell matrix
   \begin{equation}
   ^2B= \frac{1}{\sqrt{2}}\left[
   \begin{array}{cccc}
   1 & 0 & \ 0 &  e^{i\varphi} \\
   0 & 1 & \ 1 & \ 0 \\
   0 & -1 & 1 & \ 0 \\
   -e^{-i\varphi} & 0 & \ 0 & 1
   \end{array} \right]. 
\end{equation}
This braid matrix is associated to the TL algebra (see Eq.~(\ref{BTrelation}
)) with quantum dimension $d=-\alpha^2-\alpha^{-2}=\sqrt{2}$, $\alpha=e^{i3\pi/8}$, (up to an overall phase $e^{-i\pi/8}$ in comparison with Eq.~(\ref{BTrelation})
\begin{equation}\label{TLB2}
^2B=e^{-i\pi/8}(\alpha I+\alpha^{-1} \,^2T)=e^{i\pi/4} I-i \,\, ^2T.
\end{equation}

The $^2T$-matrix corresponding TL generators $^2T_i=I\otimes\cdots \underset{i,i+1}{^2T}\cdots\otimes I$ reads
\begin{equation}\label{TL42}
^2T= \tfrac{1}{\sqrt{2}}\left[
   \begin{array}{cccc}
   1 & 0 & \ 0 & \ ie^{i\varphi}\\
   0 & 1 & \ i & \ 0 \\
   0 & -i & \ 1 & \ 0 \\
   -ie^{-i\varphi} & 0 & \ 0 & \ 1
   \end{array} \right],
\end{equation}
and the quantum dimension $d=\sqrt{2}$. Substituting Eq.~(\ref{TL42}) into Eq.~(\ref{Rmu}), one obtains the Yang-Baxterized matrix, denoted as $^1\breve{R}(\theta)$ (suppose $\tan\theta=-\mu\frac{id}{2a_0}$), 

\begin{equation}\label{3Type-II}
   ^2\breve{R}(\theta,\varphi) = \left[
   \begin{array}{cccc}
   \cos\theta & 0 & \ 0 & \ \sin\theta e^{i\varphi} \\
   0 & \cos\theta & \ \sin\theta & \ 0 \\
   0 & -\sin\theta & \ \cos\theta & \ 0 \\
   -\sin\theta e^{-i\varphi} & 0 & \ 0 & \ \cos\theta
   \end{array} \right]. 
\end{equation}

\bigskip
\noindent {\bf Reducing type-II tensor YBE to lower dimension.}
 $^2T_i$ can be expressed in terms of spin-1/2 lattice sites
\begin{equation}
^2T_i=\sqrt{2}\left(|\psi_{i,i+1}\rangle\langle\psi_{i,i+1}|+|\phi_{i,i+1}\rangle\langle\phi_{i,i+1}|\right),
\end{equation}
where
\begin{equation}\label{EntangledSpin}
|\psi_{i,j}\rangle=\frac{1}{\sqrt{2}}\left(|\underset{i}{\uparrow}\underset{j}{\uparrow}\rangle-ie^{-i\varphi}|\underset{i}{\downarrow}\underset{j}{\downarrow}\rangle\right), \quad |\phi_{i,j}\rangle=\frac{1}{\sqrt{2}}\left(|\underset{i}{\uparrow}\underset{j}{\downarrow}\rangle-i|\underset{i}{\downarrow}\underset{j}{\uparrow}\rangle\right).
\end{equation}
For 4-strand T-L algebra, introducing two orthonormal basis
\begin{equation}\label{TBII}
\begin{aligned}
&|e_1\rangle=\frac{1}{\sqrt{2}}(|\psi_{12}\rangle|\psi_{34}\rangle+|\phi_{12}\rangle|\phi_{34}\rangle),\\
&|e_2\rangle=\frac{1}{\sqrt{2}}\left[(1+e^{i\varphi})|\psi_{23}\rangle|\psi_{41}\rangle-(1-e^{-i\varphi})|\phi_{23}\rangle|\phi_{41}\rangle\right]-|e_1\rangle.
\end{aligned}
\end{equation}
Graphically, 
 \begin{equation}\label{topological basis1}
\begin{split}
   &|e_1\rangle =  \frac{1}{d}\,\, \begin{tikzpicture}[baseline=-\dimexpr\fontdimen22\textfont2\relax]
\draw[gray,line width=1.0pt] (-0.1,-0.1)--(-0.1,0.4) (-0.7,-0.1)--(-0.7,0.4) (0.1,-0.1)--(0.1,0.4) (0.7,-0.1)--(0.7,0.4);
\draw[gray,line width=1.0pt] (-0.7,-0.1) arc(180:360:0.3) (0.1,-0.1) arc(180:360:0.3);
\end{tikzpicture}\, , \quad (d=\sqrt{2})\\
   &|e_2\rangle = \frac{1}{\sqrt{d^2-1}}\left(\,\, \begin{tikzpicture}[baseline=-\dimexpr\fontdimen22\textfont2\relax]
\draw[gray,line width=1.0pt] (-0.3,-0.2)--(-0.3,0.4) (-0.7,-0.3)--(-0.7,0.4) (0.3,-0.2)--(0.3,0.4) (0.7,-0.3)--(0.7,0.4);
\draw[gray,line width=1.0pt] (-0.3,-0.19) arc(240:300:0.6) (-0.7,-0.29) arc(240:300:1.4);
\end{tikzpicture} \,\,  - \frac{1}{d} \,\, \begin{tikzpicture}[baseline=-\dimexpr\fontdimen22\textfont2\relax] 
\draw[gray,line width=1.0pt] (-0.1,-0.1)--(-0.1,0.4) (-0.7,-0.1)--(-0.7,0.4) (0.1,-0.1)--(0.1,0.4) (0.7,-0.1)--(0.7,0.4);
\draw[gray,line width=1.0pt] (-0.7,-0.1) arc(180:360:0.3) (0.1,-0.1) arc(180:360:0.3);
\end{tikzpicture}\,\, \right).
\end{split}
\end{equation}
Then we have the $2\times2$ representation of TL generators for $d=\sqrt{2}$, with the $^2T_i$ matrix elements
\begin{equation}\label{TL22}
\begin{aligned}
&^2T_1|e_1\rangle=\,^2T_3|e_1\rangle=\sqrt{2}|e_1\rangle,\quad ^2T_1|e_2\rangle=\,^2T_3|e_2\rangle=0,\\
&^2T_2|e_1\rangle=\,^2T_2|e_2\rangle=\frac{1}{\sqrt{2}}\left(|e_1\rangle+|e_2\rangle\right).
\end{aligned}
\end{equation}
Then the $4\time4$ Bell braid matrix has the 2D corresponding braid matrix $^2B_i$
\begin{equation}\label{Braid1}
^2B_1=\,^2B_3=e^{-i\pi/4}\left[
   \begin{array}{cc}
   1 & 0 \\
   0 & i 
   \end{array} \right], \quad ^2B_2=\frac{1}{\sqrt{2}}\left[
   \begin{array}{cc}
   1 & -i \\
   -i & 1 
   \end{array} \right].
\end{equation}
Substituting Eq.~(\ref{TL22}) into Eq.~(\ref{Rmu})and the corresponding $2\times2$  $^1\mathcal{R}_i(\mu)$ matrices are($\tan\theta=-\mu\frac{id}{2a_0}$)
\begin{align}
&^2\mathcal{R}_1(\theta)= \,^2\mathcal{R}_3(\theta)=\left[
   \begin{array}{cc}
  e^{i\theta}& 0 \\
   0 & e^{-i\theta}
   \end{array} \right],\\
&^2\mathcal{R}_2(\theta)=\left[
   \begin{array}{cc}
   \cos\theta & i\sin\theta \\
   i\sin\theta & \cos\theta 
   \end{array} \right].
\end{align}
They satisfy the YBE and the parameters obey the Lorentz additivity,
\begin{equation}
^2\mathcal{R}_1(\theta_1)\, ^2\mathcal{R}_2(\theta_2) \,^2\mathcal{R}_1(\theta_3)=\, ^2\mathcal{R}_2(\theta_3) \,^2\mathcal{R}_1(\theta_2)\,^2\mathcal{R}_2(\theta_1). 
\end{equation}

\setcounter{equation}{0}
\renewcommand{\theequation}{B\arabic{equation}}

\section{Appendix B: Wigner $D^j$-function as the solution of Yang-Baxter equation}
Any spin coherent operator, say $D(\theta,\phi)=e^{\xi J_+-\xi* J_-}$ is identical with the Euler rotation
\begin{equation}
D(\theta,\phi)=e^{i\phi J_z}e^{i2\theta J_y}e^{-i\phi J_z},
\end{equation}
where $J_x$, $J_y$, $J_z$ are $su(2)$ operators, obeying $[J_i,J_j]=i\epsilon_{ijk}J_k$, $\epsilon_{ijk}$ the Levy-Civita symbol. $D(\theta,\phi)$ is the Wigner $D^j$-function for $(2j+1)$-D representation.
 
Then we can define the following  $D$-functions to satisfy YBE
\begin{equation}
D(\theta_1,0)D(\theta_2,\phi)D(\theta_3,0)=D(\theta_3,\phi)D(\theta_2,0)D(\theta_1,\phi),
\end{equation}
with the constraint
\begin{equation}\label{p123}
\cos\phi=\frac{1}{2}\left[\frac{(\tan\theta_1+\tan\theta_3)-\tan\theta_2}{\tan\theta_1\tan\theta_2\tan\theta_3}-1\right].
\end{equation}
Here we note that the parameter relation can be directly derived from the $su(2)$ algebraic relation, i.e. independent of the concrete representations of $D$-function. 

When $\theta_1=\theta_2=\theta_3=\theta$, YBE reduces to the braid relation, with the parameter relation
\begin{equation}
 \cos\phi=\frac{\cos2\theta}{1-\cos2\theta}.
\end{equation}

The Wigner $D$-function for spin-1/2 is
\begin{equation}
D^{1/2}=\left[\begin{array}{cc}
  \cos\theta & -\sin\theta e^{-i\phi}\\
   \sin\theta e^{i\phi} & \cos\theta
   \end{array} \right].
\end{equation}
To satisfy braid relation
\begin{equation}
D(\theta,\phi_1=0)D(\theta,\phi_2=\varphi)D(\theta,\phi_1=0)=D(\theta,\phi_2=\phi) D(\theta,\phi_1=0)D(\theta,\phi_2=\phi),
\end{equation}
the angular relation between $\theta$ and $\phi$ is
 \begin{equation}\label{ThetaPhi}
 \cos\phi=\frac{\cos2\theta}{1-\cos2\theta}.
 \end{equation} Let 
\begin{align}
&A(\theta)=VD(\theta,\phi_1=0)V^\dag=\left[\begin{array}{cc}
  e^{i\theta/2} & 0\\
  0  & e^{-i\theta/2}
   \end{array} \right],\label{A}\\
& B(\theta)=VD(\theta,\phi_2=\phi)V^\dag=\left[\begin{array}{cc}
  \cos\frac{\theta}{2}+i\sin\frac{\theta}{2}cos\phi & i\sin\phi\sin\frac{\theta}{2}\label{B}\\
   i\sin\phi\sin\frac{\theta}{2} & \cos\frac{\theta}{2}-i\sin\frac{\theta}{2}\cos\phi 
   \end{array} \right],
\end{align}
where $V=\frac{1}{\sqrt{2}}\left[\begin{array}{cc}
  1 & i \\
  i & 1
   \end{array} \right]$.
   
 \bigskip
 
   Then $A(\theta)$ and $B(\theta,\phi)$ cover the  type-I and type-II braid matrices we have mentioned above.
   \begin{itemize}
   \item Type-I braid matrix. $\theta=\pi/2, \phi=2\pi/3$.
   \begin{equation}
   A(\theta)\longrightarrow -i\left[\begin{array}{cc}
  -1 & 0 \\
  0 & 1
   \end{array} \right],\quad  B(\theta,\phi)\longrightarrow-\frac{i}{2}\left[\begin{array}{cc}
  1 & -\sqrt{3} \\
  -\sqrt{3} & -1
   \end{array} \right].
   \end{equation}
 \item Type-II braid matrix. $\theta=\pi/4, \phi=\pi/2$.
   \begin{equation}
   A(\theta)\longrightarrow e^{-i\pi/4}\left[\begin{array}{cc}
  1 & 0 \\
  0 & i
   \end{array} \right],\quad  B(\theta,\phi)\longrightarrow\frac{1}{\sqrt{2}}\left[\begin{array}{cc}
  1 & i \\
  i & 1
   \end{array} \right].
   \end{equation}
   \end{itemize}

   \setcounter{equation}{0}
\renewcommand{\theequation}{C\arabic{equation}}

   \section{Appendix C: Reduction of 3-body $S$-matrix from 8D $\breve{S}_{123}$  to 2D $\breve{S}'_{123}$}
 Usually, the  Yang-Baxter equation takes the following  form
 \begin{equation}
 \breve{R}_{12}(\theta_1)\breve{R}_{23}(\theta_2)\breve{R}_{12}(\theta_3)=\breve{R}_{23}(\theta_3)\breve{R}_{12}(\theta_2)\breve{R}_{23}(\theta_1),
 \end{equation}
 where $\breve{R}_i$ can be regarded as the 2-body $S$-matrix, and the three parameters obey a constrained relation
 \begin{equation}
 \theta_2=f(\theta_1,\theta_3).
 \end{equation} 
 
 Then the  3-body  $S$-matrix can be decomposed into 2-body $S$-matrices constrained by Yang-Baxter equation, with each 2-body $S$-matrix represented by the solution $\breve{R}$,
 \begin{equation}\label{3BodyTensor}
 \breve{S}_{123}=\breve{R}_{12}(\theta_1)\breve{R}_{23}(\theta_2)\breve{R}_{12}(\theta_3)=\breve{R}_{23}(\theta_3)\breve{R}_{12}(\theta_2)\breve{R}_{23}(\theta_1).
 \end{equation}
  
\noindent{\bf  3-body $S$-matrix in 3-qubit tensor product space.} In this paper we mainly focus on the type-II Yang-Baxter solution $\breve{R}(\theta)$ as the 2-body $S$-matrix. Substituting Eq.~(\ref{3Type-II}) into Eq.~(\ref{3BodyTensor}),  the 3-body $S$-matrix can be expressed as  
\begin{equation}\label{3MFExpS}
  \breve{S}_{123}(\eta,\beta)=e^{\eta \left(\vec{n}\cdot\vec{\Lambda}\right)},
\end{equation}
where 
\begin{eqnarray*}
 \cos\eta &=& \cos\theta_2 \cos\left(\theta_1+\theta_3\right),\\
 \sin\eta &=& \sin\theta_2\sqrt{1+\cos^2(\theta_1-\theta_3)},\\
 \vec{n} &=& \left(
  \begin{array}{ccc}
  \tfrac{1}{\sqrt{2}}\cos\beta,& \tfrac{1}{\sqrt{2}}\cos\beta,& \sin\beta
  \end{array}\right),\\
  \vec{\Lambda} 
  		&=&(-i\sigma^y\sigma^x I,\, -iI\sigma^y\sigma^x,\, -i\sigma^y\sigma^z\sigma^x),\\
  \cos\beta &=& \tfrac{\sqrt{2}\cos\left(\theta_1-\theta_3\right)}{\sqrt{1+\cos^2\left(\theta_1-\theta_3\right)}},\\               
  \sin\beta &=& \tfrac{-\sin\left(\theta_1-\theta_3\right)}{\sqrt{1+\cos^2\left(\theta_1-\theta_3\right)}}.
 \end{eqnarray*}
 Here we suppose $\varphi=0$ for simplicity, which does not influence our results. Due to the constraint of YBE, there are only two free parameters $\eta$ and $\beta$ among the three parameters $\theta_i$'s.

\medskip

\noindent{\bf  3-body $S$-matrix in anyon fusion space.} On the other hand,  due to the 2-body $\breve{R}_i$ has 2D reduction in anyon fusion bases, then the 3-body S-matrix $\breve{S}_{123}$ in tensor product space can also be reduced into a subspace named anyon fusion space spanned by $\{|e_1\rangle,\, |e_2\rangle\}$ in Eq.~(\ref{TBII}).  

For the type-II solution of YBE the reduction is ($\phi=\pi/2$ in Eq.~(\ref{B}))
\begin{equation}
\breve{R}_{12}(\theta)\longrightarrow A(\theta)=\left[\begin{array}{cc}
  e^{i\theta/2} & 0\\
  0  & e^{-i\theta/2}
   \end{array} \right],\quad \breve{R}_{23}(\theta)\longrightarrow B(\theta, \phi=\pi/2)=\left[\begin{array}{cc}
  \cos\frac{\theta}{2}& i\sin\frac{\theta}{2}\label{B}\\
   i\sin\frac{\theta}{2} & \cos\frac{\theta}{2} 
   \end{array} \right].
\end{equation}
Then the 3-body $S$-matrix in anyon fusion space should be
\begin{equation}
\begin{split}
\breve{S}'_{123}(\eta,\beta)&=A(\theta_1)B(\theta_2, \phi=\pi/2)A(\theta_3)=B(\theta_3, \phi=\pi/2)A(\theta_2)B(\theta_1, \phi=\pi/2),\\
&=\left[\begin{matrix}\cos\eta+\frac{i}{\sqrt{2}}\cos\beta\sin\eta, &  (\frac{i}{\sqrt{2}}\cos\beta+\sin\beta)\sin\eta \\ (\frac{i}{\sqrt{2}}\cos\beta-\sin\beta)\sin\eta, & \cos\eta-\frac{i}{\sqrt{2}}\cos\beta\sin\eta\end{matrix}\right].
\end{split}
\end{equation}
Where 
\begin{eqnarray}
&&\cos\eta=\cos\theta_2\cos(\theta_1+\theta_3),\\
&&\sin\eta\sin\beta=-\sin\theta_2\sin(\theta_1-\theta_3),\\
&&\cos\beta\sin\eta=\sqrt{2}\sin\theta_2\cos(\theta_1-\theta_3)=\sqrt{2}\cos\theta_2\sin(\theta_1+\theta_3).\quad \textrm{(YBE constraint)}
\end{eqnarray}
Physically, the reduced 3-body $S$-matrix $\breve{S}'_{123}$ can be regarded as describing 3-anyon interactions.

\end{document}